\documentclass[conference]{IEEEtran}
\IEEEoverridecommandlockouts
\usepackage[style=ieee,eprint=true]{biblatex}
\usepackage{tikz}
\usetikzlibrary{positioning, fit, backgrounds, arrows.meta}
\usepackage{amsmath,amssymb,amsfonts}
\usepackage{graphicx}
\usepackage{textcomp}
\usepackage{xcolor}
\usepackage{booktabs}
\usepackage{bm}
\usepackage{physics}

\def\BibTeX{{\rm B\kern-.05em{\sc i\kern-.025em b}\kern-.08em
    T\kern-.1667em\lower.7ex\hbox{E}\kern-.125emX}}

\DeclareRobustCommand{\erase}{\bgroup\markoverwith{\textcolor{red}{\rule[.5ex]{2pt}{0.4pt}}}\ULon}

\begin{document}

\title{Solving Capacitated Vehicle Routing Problem with Quantum Alternating Operator Ansatz and Column Generation
    \thanks{
    This work was performed for the Council for Science, Technology and Innovation (CSTI), Cross-ministerial Strategic Innovation Promotion Program (SIP), "Promoting the application of advanced quantum technology platforms to social issues" (Funding agency: QST).
  }
}

\author{
  \IEEEauthorblockN{
    Wei-hao Huang\IEEEauthorrefmark{1},
    Hiromichi Matsuyama\IEEEauthorrefmark{1},
    and Yu Yamashiro\IEEEauthorrefmark{1}
  }
  \IEEEauthorblockA{
    \IEEEauthorrefmark{1}\textit{Jij Inc.}, 3-3-6 Shibaura, Minato-ku, Tokyo, 108-0023, Japan
  }
  \IEEEauthorblockA{
    w.huang@j-ij.com,
    h.matsuyama@j-ij.com,
    y.yamashiro@j-ij.com,
  }
}

\maketitle
\begin{abstract}
This study proposes a hybrid quantum-classical approach to solving the Capacitated Vehicle Routing Problem (CVRP) by integrating the Column Generation (CG) method with the Quantum Alternating Operator Ansatz (QAOAnsatz). 
The CG method divides the CVRP into the reduced master problem, which finds the best combination of the routes under the route set, and one or more subproblems, which generate the routes that would be beneficial to add to the route set.
This method is iteratively refined by adding new routes identified via subproblems and continues until no improving route can be found.
We leverage the QAOAnsatz to solve the subproblems.
Our algorithm restricts the search space by designing the QAOAnsatz mixer Hamiltonian to enforce one-hot constraints.
Moreover, to handle capacity constraints in QAOAnsatz, we employ an Augmented Lagrangian-inspired method that obviates the need for additional slack variables, reducing the required number of qubits.
Experimental results on small-scale CVRP instances (up to 6 customers) show that QAOAnsatz converges more quickly to optimal routes than the standard QAOA approach, demonstrating the potential of this hybrid framework in tackling real-world logistical optimization problems on near-term quantum hardware.
\end{abstract}


\section{Introduction}
Combinatorial optimization focuses on finding solutions that minimize or maximize a given cost function while satisfying the constraints~\cite{korte2011combinatorial, Laurence_Wolsey2020-gj, conforti2014integer}. 
There are numerous optimization problems in the industrial field.
Planning the delivery route is one of the crucial tasks in the logistics. 
Capacitated Vehicle Routing Problem (CVRP) is one of the simplest routing problems for optimizing the delivery of goods with capacity constraints to minimize delivery costs~\cite{dantzig1959truck, ralphs2003capacitated}.
Various algorithms for solving CVRP have been proposed~\cite{baldacci2012recent, costa2019exact}.

In parallel, quantum computing, using the principles of quantum mechanics, has the potential speedups for specific problems, such as integer factorization~\cite{shor1994algorithms}, matrix computation~\cite{wan2018asymptotic} and simulating the classical dynamics~\cite{babbush2023exponential}.
These advantages lead to potential applications for quantum computers.
However, quantum computing faces formidable challenges in the Noisy Intermediate Scale Quantum (NISQ) era. 
Due to the number of qubits, the quality of qubits and gate operation present significant hurdles that demand our attention and innovation.
These constraints currently impede the application of these algorithms on existing small-size quantum devices.
In the optimization field, the Quantum Approximate Optimization Algorithm (QAOA)~\cite{farhi2014quantum} and its generalization is one of the most famous NISQ algorithms. 
QAOA successfully solves combinatorial optimization problems without the constraints, such as the max-cut problem~\cite{wang2018quantum, herrman2021impact}.
However, QAOA is challenged in handling constraints, as it can deal with them only through relaxation, even though most real-world problems incorporate various types of constraints.
To overcome this obstacle, the Quantum Alternating Operator Ansatz (QAOAnsatz)~\cite{Hadfield2019-lb} was introduced.
QAOAnsatz is suitable for optimization problems with constraints that always need to be satisfied, such as graph-coloring~\cite{wang2020xy} and the maximum independent set~\cite{saleem2020max}.

Many of the QAOA studies have dealt with relatively simple problems.
However, in recent years, there has been much interest in how to handle problems that are closer to real-world problems, such as VRP or CVRP, using quantum computing~\cite{feld2019hybrid,palackal2023quantum, Weinberg2023-ka, azad2022solving, Xie2024-sb}.
There are two types of methods for solving CVRP using quantum algorithms.
The first method is the direct method, which solves the entire problem using a quantum algorithm~\cite{Xie2024-sb}.
The advantage of the direct method is that it can maintain the optimality of the original problem.
However, the problem becomes more complicated and requires many qubits to express the entire problem.
The other method is the decomposition method, which divides the CVRP into smaller subproblems and solves only part using a quantum algorithm~\cite{feld2019hybrid, palackal2023quantum, Weinberg2023-ka}.
The advantage of the decomposition method is that it can reduce the difficulty of handling the problem and the number of required qubits. 
However, most of them are heuristics, and the quality of the solution depends not only on the quality of the classical and quantum algorithms but also on how they are combined.

Recently, the Column Generation (CG) method has attracted attention in the field of application of quantum optimization~\cite{hirama2023efficient, da2023quantum} and classical QUBO solver~\cite{kanai2024annealing}.
Our study proposes a hybrid approach combining QAOAnsatz with CG method~\cite{desaulniers2006column} to address CVRP.
CG method is the decomposition method, which devices the CVRP into the restricted master problem and subproblems.
The restricted master problem finds the best combination of the routes under the route set.
Here, the subproblems generate the routes that would be beneficial to add to the route set.
Using the CG method makes CVRP easier to handle than when generating the routes for all vehicles simultaneously.
Moreover, the CG method can check how far from optimal solutions by checking the objective value of the subproblems.
However, the subproblem is still a combinatorial optimization problem.
Thus, we applied QAOAnsatz to solve the subproblems in our hybrid CG method.

Dealing with the capacity constraint of the vehicles is one of the challenges in the quantum approach to CVRP.
In this study, we have applied an Augmented Lagrangian-inspired Method (ALiM) to deal with the capacity constraint~\cite{Barrera2024unbalanced, cellini2024qal}.
The ALiM reduces the required qubits when the problem contains inequality constraints compared to when the slack variables are introduced.

The paper will be structured as follows: Firstly, Section ~\ref{sec:Background} will discuss the QAOA and QAOAnsatz, a fundamental component of this study, and then we will explain the Quadratic Unconstrained Binary Optimization, which QAOA can deal with.
In Sec.~\ref{sec:Methods}, we will explain the CVRP, its mathematical form for the CG method, and the CG's routine. 
We will present a new approach to solving subproblems using the QAOAnsatz, reducing the required qubits.
We will present the experiment results obtained using our method for the problem in Sec.~\ref{sec:results}, demonstrating the practicality and effectiveness of our approach.
Finally, in Sec.~\ref{sec:summary}, we summarize our results and discuss future works.

\section{Background}
\label{sec:Background}

\subsection{Quantum Approximate Optimization Algorithm}
The Quantum Approximate Optimization Algorithm (QAOA) is a quantum-classical variational algorithm used to solve the Quadratic Unconstrained Binary Optimization (QUBO) problem~\cite{farhi2014quantum}.
The approach begins by converting an objective function~$C(\bm{x})$, into a corresponding cost Hamiltonian $H_C$, such that $H_C|\bm{x}\rangle = C(\bm{x})|\bm{x}\rangle$.
Because we can map QUBO problems to the problem of finding the ground state of an Ising Hamiltonian, $H_C$ is commonly constructed in the form of an Ising Hamiltonian.
In addition to $H_C$, QAOA employs a mixer Hamiltonian $H_M = \sum_i X_i$, which helps explore the solution space.
The algorithm applies alternating evolution sequences under $H_C$ and $H_M$, over $p$ rounds, starting from an initial state $|\psi_0\rangle$.
A common choice for $|\psi_0\rangle$ is the equal superposition of all $n$-qubits computational basis states $|\psi_0\rangle = (|0\rangle + |1\rangle)^{\otimes n}/\sqrt{2^n}$, ensuring an unbiased starting point. 
This iterative application of two operators prepares the following state: 
\begin{equation}
|\psi(\bm{\gamma},\bm{\beta})\rangle = e^{-i\beta_p H_M}e^{-i\gamma_p H_C} \cdots e^{-i\beta_{1} H_M}e^{-i\gamma_{1} H_C} |\psi_0\rangle,
\label{eq:qaoa}
\end{equation}
where the $2p$ parameters $\bm{\gamma} = (\gamma_1, \ldots, \gamma_p)$ and $\bm{\beta} = (\beta_1, \ldots, \beta_p)$ are optimized via classical optimization techniques.
The goal is to minimize the expectation value $ \langle\psi(\bm{\gamma},\bm{\beta})|H_C|\psi(\bm{\gamma},\bm{\beta})\rangle $.
After determining the optimal parameters $\bm{\gamma}^*$ and $\bm{\beta}^*$, the final step is to sample from the optimized state $|\psi(\bm{\gamma}^*,\bm{\beta}^*)\rangle$ to obtain the optimal solution.

\subsection{Quantum Alternating Operator Ansatzes}
Quantum Alternating Operator Ansatzes (QAOAnsatz) is the generalized QAOA framework.
QAOA framework is unique in designing problem-specific mixing operators that search directly within the feasible region of a given problem.
Rather than employing a simple $X$-mixer as in standard QAOA, QAOAnsatz allows for more flexible operator choices.
In this framework, Eq.~\eqref{eq:qaoa} is extended to:
\begin{equation}
\ket{\psi(\bm{\gamma},\bm{\beta})} = U_M(\beta_p)U_P(\gamma_p) \cdots U_M(\beta_1)U_P(\gamma_1) \ket{\psi_0},
\label{eq:qaoansatz}
\end{equation}
where $U_M(\beta)$ is a mixing operator, and $U_P(\gamma)$ is a phase-separation operator.
The mixing operator, $U_M(\beta)$, is designed by the structure of the feasible set, and the phase-separation operator, $U_P(\gamma)$, is created by the cost function $ C(x) $.
The circuit construction depends on the specific problem; the detailed design for our problem is presented in Section~\ref{sec:Methods}.
By ensuring the initial state $\ket{\psi_0}$ is a feasible solution or a superposition of feasible solutions, QAOAnsatz restricts the search space to the feasible region.
These limitations sure potentially leads to a more efficient search process than QAOA.

\subsection{Quadratic Unconstrained Binary Optimization \label{sec:qubo}}
Many real-world optimization problems can be formulated as constrained optimization problems.
Since QAOA only addresses QUBO problems, a translation step is necessary.
Consider a constrained binary optimization problem with $m$ linear constraints formulated as:
\begin{equation}
\label{Eq:qbo}
\begin{split}
    \min_{x \in \{0,1\}^n} 
    &\sum_{i,j}Q_{ij}x_ix_j\\
    \mathrm{s.t.}\quad  
    &\sum_{i}A_{ki}x_i \leq b_k\quad \forall k\in \{1,\dots,m\},\\
    &x_i \in \{0,1\}
\end{split}
\end{equation}
where $\bm{x}$ are binary decision variables and $Q \in \mathbb{R}^{n \times n}$ is the QUBO matrix.
In this formulation, $A \in \mathbb{R}^{m \times n}$ and $\bm{b} \in \mathbb{R}^m$ are the constant matrix and vector related to the constraints.
A commonly used method is the penalty method~\cite{lucas2014ising}.
We apply this approach, then the constrained optimization problem in Eq.~\eqref{Eq:qbo} can be formulated as a QUBO problem:
\begin{equation}
\min_{x \in \{0,1\}^n} \sum_{i,j}Q_{ij}x_ix_j + \sum_{k} \mu_k \qty(\sum_iA_{k i}x_i + s_{k} - b_{k} )^2.
\label{eq:penalty}
\end{equation}
where $\mu_k\geq 0$ is a penalty coefficient that ensures the equality constraint is satisfied in the QUBO formulation.
The $s_{k}$ is a slack variable introduced to convert the inequality constraint into an equality constraint. 

Introducing slack variables into the problem requires additional qubits.
Assuming $m$ slack variables, each ranging from $0$ to $b$, and applying a logarithmic encoding to convert the integer values into binary variables, the number of required qubits is $m\lceil\log_2 b\rceil$.
Consequently, the number of qubits increases with the number of inequality constraints.
Moreover, introducing slack variables into the problem fundamentally depends on the value of $A_{ki}$ and $b_k$. 
If either $A_{ki}$ and $b_k$ contain irrational values, then the slack variables should not be treated as discrete integers but rather as real values.
In such cases, encoding $s_k$ directly as qubits may not be feasible without approximation.
In the case that both $A_{ki}$ and $b_k$ take only the rational values, we can deal with the slack variable as the integer variable by multiplying the least common multiple of the denominator of $A_{ki}$ and $b_k$.
However, this multiplication changes the energy scale of the penalty term.
Thus, tuning $\mu_k$ becomes more difficult.
\begin{figure*}[h!]
\centering
\includegraphics[width=\linewidth]{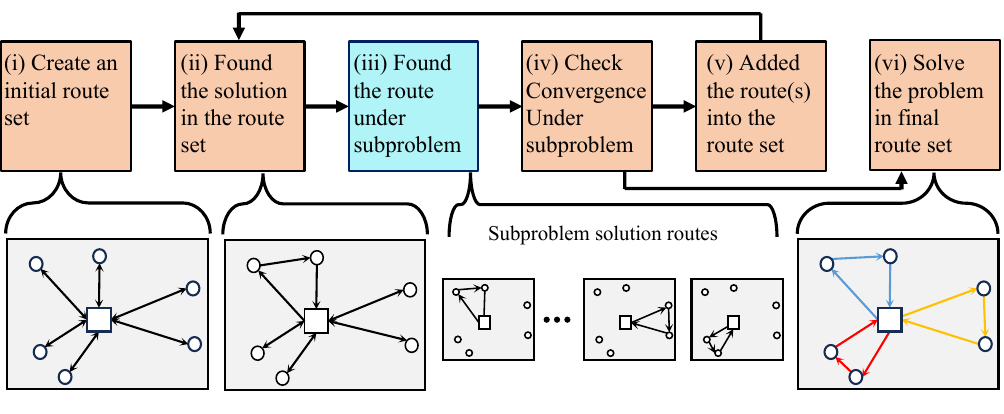}
\caption{
Illustration of the CG method applied to CVRP. 
The background color indicates which devices are used at each step.
Orange corresponds to the classical device, and light blue corresponds to the quantum device.
(i) Start with an initial route set.
(ii) Solve the restricted master problem current route set. 
(iii) Identify new, potentially cost-reducing routes via solving the subproblem by the quantum devices.
(iv) Check the algorithm is converged.
(v) If it is not converged, add the newly found routes to the route set.
(vi) Once no more improvements are found, solve the master problem with the final set of routes to obtain the optimal solution.
Examples of subproblem-generated routes (in black) are shown at the bottom, and the final solution is shown in color, which is iteratively incorporated back into the master problem.}
\label{fig:big_picture}
\end{figure*}

\section{Methods\label{sec:Methods}}
In this section, we first introduce the CVRP, highlighting its fundamental constraints and the combinatorial complexities that arise when determining delivery routes under capacity limitations. 
We then present the column generation (CG) algorithm as a systematic method to overcome these complexities by iteratively refining a dual problem of the restricted master problem (RMP) through subproblem (SP) solutions.
Finally, we discuss our quantum-based subproblem formulation.
The discussed methods, such as the Augmented Lagrangian-inspired Method and QAOAnsatz, are employed to handle capacity constraints and route selection within a quantum computing framework.

\subsection{Capacitated Vehicle Routing Problems}
Capacitated Vehicle Routing Problems (CVRP) are complex combinatorial optimization problems~\cite{dantzig1959truck} that involve determining the delivery routes of multiple vehicles serving a set of customers.
Each vehicle has a limited carrying capacity, and the objective is to minimize the total distance traveled by all vehicles.
The CVRP has the following constraints on the solution routes:
\begin{enumerate}
\item Every customer is visited exactly once by a vehicle.
\item The total demand of customers on each route does not exceed the carrying capacity of the assigned vehicle.
\item Each route begins and ends at the depot.
\end{enumerate}
A major challenge in solving the CVRP is the combinatorial explosion in the number of potential routes.
CVRP is similar to the Travelling Salesman problem (TSP).
However, since CVRP allows multiple vehicles, it is necessary to consider routes that visit all customers and routes that visit only some customers.
Therefore, the number of potential routes increases significantly, even for the same problem size.

Although the set of all possible routes is extensive, only a relatively small subset of these routes is typically relevant to constructing an optimal or near-optimal solution. 
Techniques such as CG are employed to identify and utilize these meaningful routes efficiently.

\subsection{Column Generation Algorithm}

The essence of the column generation (CG) method is to decompose the original problem into a Restricted Master Problem (RMP) and one or more Subproblems (SPs)~\cite{desaulniers2006column}.
The algorithm then iteratively refines the solution by solving these problems in sequence.
In the CVRP context, the RMP begins with a limited set of routes, and the SPs are tasked with identifying new, potentially cost-reducing routes that can be added to it.

Figure~\ref{fig:big_picture} provides a flow chart and schematic picture for the CG method applied to CVRP.
The process can be summarized as follows: (i) The process begins with creating an initial route set, which provides a feasible but not optimal solution to the RMP. 
This step serves as the starting point for the iterative optimization process.
(ii) Then, we use the current route set as input, and the RMP is solved to determine the best combination of the routes for the given set of routes.
This solution is temporary and may still leave room for improvement.
(iii) Based on the current solution of RMP, SPs are constructed to identify new routes to reduce the overall cost further. 
(iv) We perform the convergence check based on the objective value of SPs.
If the algorithm converges, we skip step~(vi) directly. 
Otherwise, we move to step~(v).
(v) The cost-reducing new routes generated by solving the SPs are added to the current route set.
The number of the input routes $K$ is one of the hyperparameters of this algorithm.
We will discuss the influence of $K$ in Sec.~\ref{sec:results}.
After step~(v), we return to step~(ii) and resolve the RMP with the updated route set to ensure iterative improvement.
(vi) The final route set is used to solve the RMP one last time to generate the final solution for the CVRP.

\subsubsection{Restricted Master Problem for CVRP}

To introduce the RMP for the CVRP, we first consider the master problem. 
In the master problem, instead of constructing the routes in advance, we assume access to a complete set $\tilde{\mathcal{R}}$ containing all feasible routes.
By using this route set, we can formulate the CVRP as follows:
\begin{equation}
    \begin{split}
        & \min\quad \sum_{r \in \tilde{\mathcal{R}}} d_r x_r \\ 
    &\text{s.t.}  \quad \sum_{r \in \tilde{\mathcal{R}}} a_{ri} x_r = 1, \quad \forall i \in \mathcal{C}, \\
    &\quad\ \quad x_r \in \{0, 1\} \quad \forall r \in \tilde{\mathcal{R}},
    \end{split}
    \label{eq:master}
\end{equation}
where $x_r$ is the binary decision variable defined as:
\begin{equation}
x_r=\left\{
\begin{array}{ll}
1 & \mathrm{if}\ r \in \tilde{\mathcal{R}} \ \mathrm{is\ selected}\
\\0 & \mathrm{otherwise}
\end{array}\right. .
\end{equation}

Here, $d_r$ denotes the total distance of route $r$, and $a_{ri}$ is a binary parameter equal to $1$ if customer $i$ is included in route $r$, and $0$ otherwise. 
The total number of customers, including the depot, is $N$, and we assume the depot to be customer $0$, so the customer set is defined as $\mathcal{C}=\{1,\dots, N-1 \}$. 
This formulation aims to minimize the total distance traveled while ensuring that each customer is served exactly once.

In the master problem, Eq.~\eqref{eq:master}, we assumed access to the complete set of all feasible routes $\tilde{\mathcal{R}}$.
However, this assumption is only feasible for the smallest problem instances.
We apply the CG method to address this issue.
Within CG, we begin with a limited initial subset of feasible routes $\mathcal{R} \subset \tilde{\mathcal{R}}$.
Then, we define the following RMP:
\begin{equation}
    \begin{split}
        & \min\quad \sum_{r \in \mathcal{R}} d_r x_r \\ 
    &\text{s.t.}  \quad \sum_{r \in \mathcal{R}} a_{ri} x_r = 1, \quad \forall i \in \mathcal{C}, \\
    &\quad\ \quad x_r \in \{0, 1\} \quad \forall r \in \mathcal{R}.
    \end{split}
    \label{eq:RMP}
\end{equation}

Next, we consider the linear relaxation of the RMP. 
Specifically, we replace the binary variable $x_r$ with a continuous variable $x'_r$.
Here, we modify the equality constraint~$\sum_{r \in \mathcal{R}} a_{ri} x_r = 1$ to an inequality constraint~$\sum_{r \in \mathcal{R}} a_{ri} x'_r \geq 1$, which will not change the optimal solution of the linear relaxation of the Eq.~\eqref{eq:RMP}.
The linear relaxation of the RMP is then defined as:
\begin{equation}
\begin{split}
& \min\quad \sum_{r \in \mathcal{R}} d_r x'_r \\
&\text{s.t.}\quad \sum_{r \in \mathcal{R}} a_{ri} x'_r \geq 1, \quad \forall i \in \mathcal{C}, \\
&\quad \quad x'_r \geq 0, \quad \forall r \in \mathcal{R}.
\end{split}
\label{eq:RMP_linear}
\end{equation}

To derive the dual problem of this linear relaxation, we introduce nonnegative dual variables $y_i\ge 0$ for each constraint.
We obtain the inequality:
\begin{equation}
\sum_{r\in\mathcal{R}}\left(\sum_{i\in\mathcal{C}}a_{ri}y_i\right) x'_r \geq \sum_{i\in\mathcal{C}}y_i.
\end{equation}
Then, by assuming 
\begin{equation}
d_r \geq
\sum_{i \in \mathcal{C}} a_{ri}\,y_i,
\quad \forall r \in \mathcal{R},
\label{eq:dual_constraint}
\end{equation}
we obtain the following relationship:
\begin{equation}
\sum_{r \in \mathcal{R}} d_r x'_r \geq  \sum_{r\in\mathcal{R}}\left(\sum_{i\in\mathcal{C}}a_{ri}y_i\right) x'_r \geq \sum_{i\in\mathcal{C}}y_i.
\label{dual inequality}
\end{equation}
By Eq.~\eqref{dual inequality}, any dual solution $\bm{y}$ which satisfies Eq.~\eqref{eq:dual_constraint} provides a lower bound on the linear relaxation of the RMP's objective. 
Consequently, the dual problem is defined as the maximization of this lower bound,
\begin{equation}
\begin{split}
& \max_y \sum_{i \in \mathcal{C}} y_i \\
\text{s.t.} & \sum_{i \in \mathcal{C}} a_{ri}y_i \leq d_r, \quad \forall r \in \mathcal{R}, \\
&  y_i \geq 0, \quad \forall i \in \mathcal{C}.
\end{split}
\label{eq:dual_RMP}
\end{equation}
Here, $y_i$ can be interpreted as the “profit” for covering customer $i$, and the objective $\sum_{i\in\mathcal{C}} y_i$ is maximized under the route constraints.
Once the dual problem is solved, the optimal dual solution $y_i^*$ is then used in the subproblem (column-generation step) to guide the selection or creation of new routes to be added into $\mathcal{R}$.

\subsubsection{The Subproblem for CVRP \label{subsubsec:sp}}
In the CVRP, the subproblem identifies new feasible routes that can potentially reduce the overall transportation cost.
Formally, the subproblem is formulated as:
\begin{equation}
\label{Eq:sub}
\begin{aligned}
\underset{x}{\text{min}}
&\sum_{i=0}^{N-1} \sum_{j=0}^{N-1} \sum_{t=0}^{T-1} d_{i,j} x_{i,t} x_{j,(t+1)\!\!\!\!\!\mod\!T} -\sum_{t=0}^{T-1} \sum_{i=0}^{N-1} x_{i,t}  y_{i}^* \\
\text{s.t.}
&\sum_{t=0}^{T-1} \sum_{i=0}^{N-1} w_{i} x_{i,t} \leq W, \quad \forall t \in \{0, \dots, T-1\}, \\
&\sum_{i=0}^{N-1} x_{i,t} = 1, \quad \forall i \in \{1, \dots, T-1\}, \quad  \\
&\sum_{t=0}^{T-1} x_{i_c,t} \leq 1, \quad \forall i_c \in \mathcal{C}, \quad \\
&x_{0,0} = 1,\\
&x_{i_c,0} = 0\quad \forall i_c \in \mathcal{C}, \\
&x_{i,t} \in \{0,1\},
\end{aligned}
\end{equation}
where $W$ denotes the vehicle's capacity, $d_{i,j}$ represents the distance or traveling cost from customer $i$ to $j$, $w_i$ is the demand of customer $i$, and $x_{i,t}$ is a binary decision variable that equals 1 if the vehicle is at customer $i$ at time $t$ and 0 otherwise.
The length of time is divided into discrete periods, denoted by $T$.

This subproblem shares similarities with the quadratic formulation of the Traveling Salesman Problem~\cite{lucas2014ising}.
But they have three key differences.
First, instead of minimizing the total distance, the objective function focuses on minimizing the reduced cost, which determines whether a new route should be introduced to enhance the solution. 
The reduced cost refer to direct travel costs and incorporates the dual values of constraints.
The reduced cost reflects the benefit of modifying them. 
Second, we enforce inequality constraints to ensure that the vehicle's capacity is not exceeded.
Third, unlike the TSP, not all customers must be visited in a single tour, implying that the route length does not necessarily equal the total number of customers $N$.
Instead, we introduce a parameter $T (\leq N)$ to represent the number of time steps, allowing the depot to be visited multiple times.

In the CG method, the dual problem of the RMP and subproblems are solved iteratively.
At the $k$th iteration step, the dual problem of the RMP is solved with a limited subset of routes $\mathcal{R}_k$ to obtain the dual values $y^*$ from the dual formulation (Eq.~\eqref{eq:dual_RMP}).
These dual values are subsequently used in subproblem (Eq.~\eqref{Eq:sub}) to identify new routes with negative reduced costs.
The feasible solutions of the subproblem, $\tilde{\bm{x}}$, form the route $r$.
Thus, we can rewrite the cost function of the subproblem as the reduced cost of a candidate route $r$:
\begin{equation}
\label{Eq:cg}
\begin{aligned}
\bar{c}_r = d_r - \sum_{i \in \mathcal{C}} a_{ri} y_i^*.
\end{aligned}
\end{equation}
Here, $d_r$ is the total distance of route $r$ which is calculated by $ d_r = \sum_{j=0}^{N-1} \sum_{t=0}^{T-1} d_{i,j} \tilde{x}_{i,t} \tilde{x}_{j,(t+1)\!\!\!\!\!\mod\!T}$ and $a_{ri}=\sum_{t=0}^{T-1} x_{i,t}$, which indicates whether customer $i$ is visited in route $r$.
If a route $r^*$ is found such that $\bar{c}_{r^*} < 0$, it is added to the current route set $\mathcal{R}_k$, forming an updated set $\mathcal{R}_{k+1}$.
This process repeats iteratively, ensuring that new routes are continually introduced into the route set $R$.
The loop terminates when no more routes with negative reduced costs are found, satisfying:
\begin{equation}
\label{Eq:cg2}
\begin{aligned}
\min_{r \in \mathcal{R}_k} \bar{c}_r \geq 0.
\end{aligned}
\end{equation}
This condition is based on the constraint in the dual problem of the RMP.
At this point, the solution to the RMP over the final route set $\mathcal{R}^*$ is optimal.
This iterative process provides an efficient exploration of the solution space, progressively refining the solution to minimize the overall transportation cost.

\subsection{Subproblem Formulation for Quantum Algorithm}
The subproblem has the same form as Eq.~\eqref{Eq:qbo}.
We can deal with this problem by applying QAOA with penalty-based reformulation, which we explain in Sec.~\ref{sec:qubo}.
However, straightforward penalty-based reformulations often lead to a higher number of qubits and difficulty dealing with the constraints, so alternative methods are needed to address these problems efficiently.

\subsubsection{Augmented Lagrangian-inspired Method}
As described in Sec.~\ref{sec:Background}, a QUBO problem formulation (Eq.~\eqref{eq:penalty}) often needs a penalty method for the conversion.
However, this method requires more qubits because it necessitates converting inequality constraints into equality constraints by adding the slack variables.
In this study, we use a method inspired by the Augmented Lagrangian Method (ALM) to generate QUBO~\cite{Barrera2024unbalanced, cellini2024qal}.
In this paper, we call this method an Augmented Lagrangian-inspired Method (ALiM).
ALiM converts Eq.~\eqref{Eq:qbo} into the following form:
\begin{equation}
\begin{aligned}
\min_x\sum_{i,j}Q_{ij}x_ix_j &+ \sum_{k} \lambda_k \qty(\sum_{i} A_{k i}x_i  - b_{k} ) \\
&+\sum_{k} \mu_k \qty(\sum_{i}A_{k i}x_i  - b_{k} )^2 ,
\label{eq:ALM}
\end{aligned}
\end{equation}
where the $\lambda_j$ are Lagrange multipliers, and $\mu_i \geq 0$ are the penalty parameters. 
Equation~\eqref{eq:ALM} is still in a QUBO form.
The advantage of ALiM is that it eliminates the need for slack variables compared to the penalty method.
As a result, it is possible to reduce the number of qubits required, although we need to tune the parameters for linear terms. 
We note that our different handling of inequality constraints, compared to the textbook notation of the Augmented Lagrangian Method~\cite{birgin2014practical}, is due to the inability to use expressions such as $\max$ on quantum devices.

In this study, we set $\mu_k = \lambda_k$ for each constraint group for simplicity.
By using ALiM, we set $\lambda_1$ for the capacity constraint $\sum_{t=0}^{T-1} \sum_{i=0}^{N-1} w_{i} x_{i,t} \leq W$ becomes
\begin{equation}
    \lambda_1\left( \sum_{t=0}^{T-1} \sum_{i=0}^{N-1} w_{i} x_{i,t} - W \right) + \lambda_1\left( \sum_{t=0}^{T-1} \sum_{i=0}^{N-1} w_{i} x_{i,t} - W \right)^2.
\label{eq:capacity_alim}
\end{equation}
Moreover, the we set $\lambda_2$ for the constraint $\sum_{i=0}^{N-1} x_{i,t} \leq 1$ becomes
\begin{equation}
\begin{aligned}
&\lambda_2\sum_{t=0}^{T-1}\left( \sum_{i=0}^{N-1} x_{i,t} - 1 \right) + \lambda_2\sum_{t=0}^{T-1}\left( \sum_{i=0}^{N-1} x_{i,t} - 1 \right)^2  \\
=& \lambda_2\sum_{t=0}^{T-1}  \sum_{i=0}^{N-1} x_{i,t} \left( \sum_{i=0}^{N-1} x_{i,t} - 1 \right) .
\label{eq:onetime_alim}
\end{aligned}
\end{equation}
Equation~\eqref{eq:onetime_alim} becomes a well-known quadratic penalty.
We relaxed Eq.~\eqref{Eq:sub} by adding Eq.~\eqref{eq:capacity_alim} and Eq.~\eqref{eq:onetime_alim} to the objective function.
Thus, after applying ALiM, the subproblem becomes:
\begin{equation}
\begin{split}
\min_x\
& \sum_{i=0}^{N-1} \sum_{j=0}^{N-1} \sum_{t=0}^{T-1} d_{i,j} x_{i,t} x_{j,(t+1)\!\!\!\!\!\!\mod\!T} - \sum_{t=0}^{T-1} \sum_{i=0}^{N-1} x_{i,t}  y_{i}^* \\
 +& \lambda_1 \left( \sum_{t=0}^{T-1}\sum_{i=0}^{N-1} w_{i} x_{i,t} - W \right)  \\
 +& \lambda_1 \left( \sum_{t=0}^{T-1}\sum_{i=0}^{N-1} w_{i} x_{i,t} - W \right)^2 \\
 +& \lambda_2 \sum_{t=0}^{T-1} \sum_{i=0}^{N-1} x_{i,t} \left( \sum_{i=0}^{N-1} x_{i,t} - 1 \right) . \\
\text{s.t.}\quad 
& \sum_{i=0}^{N-1} x_{i,t} = 1, \quad \forall t \in \{1, \dots, T-1\}, \quad \\
& x_{0,0} = 1\\
& x_{i_c,0} = 0, \quad \forall i_c \in \mathcal{C}, \\
& x_{i,t} \in \{0,1\}.
\label{eq:total_alim}
\end{split}
\end{equation}

Compared to a simple slack variable method, ALiM generally requires fewer qubits because it embeds constraints directly into the objective function via penalty terms rather than relying on separate slack variables. 
In the slack variable method, each constraint typically needs its own auxiliary variable to ensure feasibility.
Equation~\eqref{Eq:sub} has one capacity constraint and $N-1$ time constraints.
The capacity constraint generates $\lceil \log_2 W  \rceil$ binary variables, and each time constraint requires one binary variable.
As a result, $N \times (T+1) + \lceil \log_2 W  \rceil -1$ qubits are required for the slack variable approach.
In contrast, ALiM only requires $N \times T$ qubits.

\subsubsection{QAOAnsatz}
\label{QAOAnsatz}
There still exists a one-hot equality constraint in Eq.~\eqref{eq:total_alim}.
In this study, we apply the QAOAnsatz to this constraint.
The core idea of the QAOAnsatz is to design the mixer Hamiltonian to search for the feasible space of the problem.
The design of the mixer Hamiltonian for this one-hot type constraint is discussed in~\cite{wang2020xy}.
We can use the $XY$-mixer Hamiltonian:
\begin{equation}
\begin{split}
H_M = \frac{1}{2}\sum^{N-1}_{i=0}\sum^{T-1}_{t=0}{\left(X_{i,t} X_{(i+1) \!\!\!\!\!\!\mod\!N,t} + Y_{i,t} Y_{(i+1) \!\!\!\!\!\!\mod\!N,t} \right)}.
\end{split}
\end{equation}
This is appropriate for the one-hot type constraint.
The form of $U_M(\beta)$ in Eq.~\eqref{eq:qaoansatz} is $U_M(\beta) = \exp\left( -i\beta H_M \right)$.
Since the mixer Hamiltonian only searches the space that satisfies the one hot constraint, the initial state $\ket{\psi_0}$ also needs to satisfy the one hot constraint. 
In this study, we used a single solution that satisfies a single onehot constraint as the initial state.
Since we incorporated the one-hot type constraint into the mixer Hamiltonian $H_M$, Eq.~\eqref{eq:total_alim} becomes a QUBO problem.
The cost Hamiltonian $H_C$ can be obtained by simply converting this QUBO into an Ising Hamiltonian.
Then, $U_P(\gamma)$ in Eq.~\eqref{eq:qaoansatz} is $U_P(\gamma) = \exp\left( -i\gamma H_C \right)$.

\begin{figure}[t]
\centering
\includegraphics[width=\linewidth]{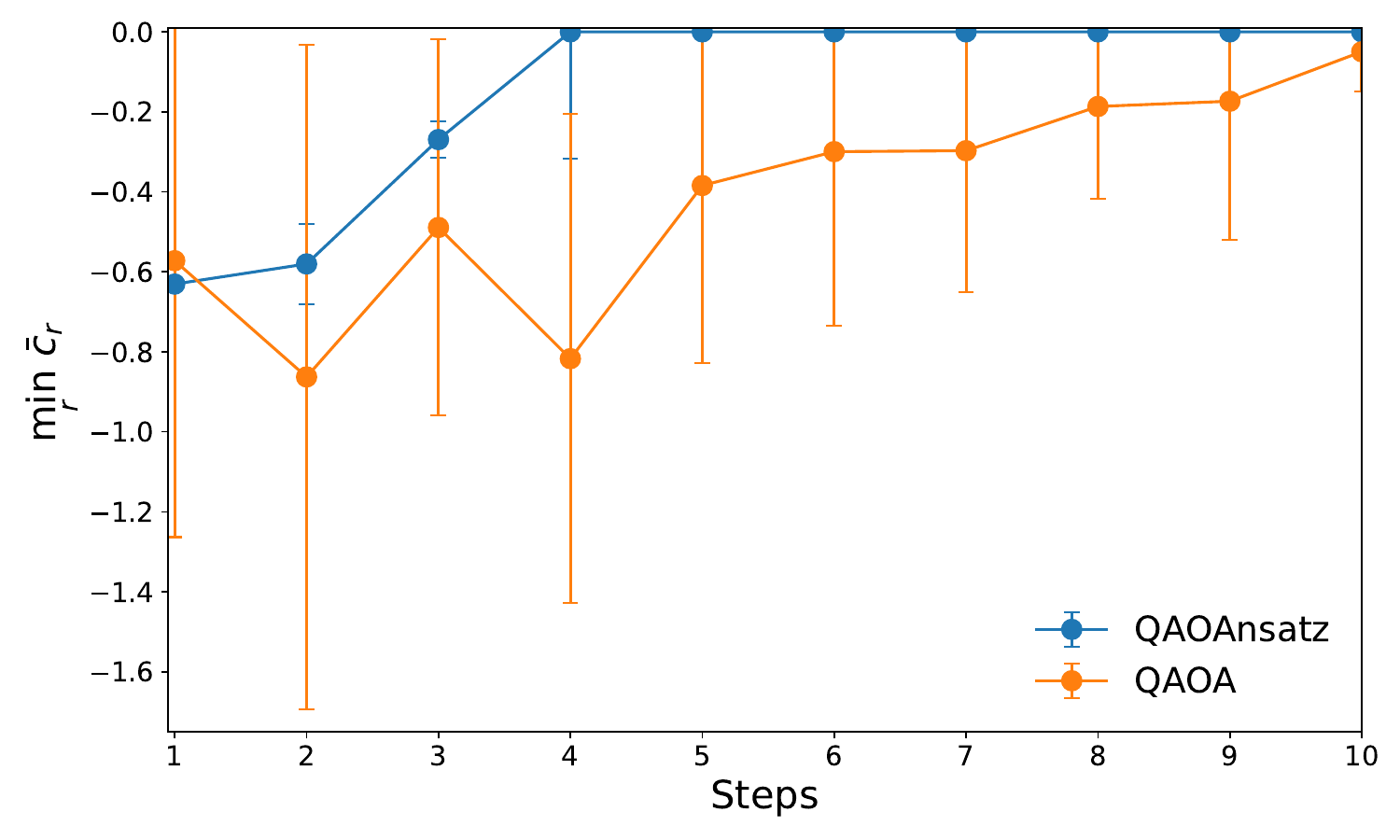}
\caption{
 The comparison between the subproblem objective values, $\min_r\bar{c}_r$, when using the QAOA and QAOAnsatz.
 In this plot, we have used four customer problems.
 QAOAnsatz shows faster convergence than QAOA.
}
\label{fig:compare}
\end{figure}

\section{Experimental Result}
\label{sec:results}
This study addresses the CVRP for randomly generated instances with four to six customers.
In all our instances, the customer's positions are randomly distributed in the $[0,1] \times [0,1]$, and the depot is at the center of this regime.
The vehicle capacity is fixed at $W = 25$, and the customer demands $w_i$ drawn randomly from the range $[1, 15]$.
In our experiments, we choose relatively smaller $W$ than the total demands because if $W$ is larger than the total demand, the CVRP becomes the same as TSP.

The CG method requires the initial route set.
For simplicity, we initialize the route set by connecting each customer directly to the depot.
This approach provides a straightforward baseline while ensuring that the constraints on vehicle capacity and route structure are satisfied.
These configurations form the CVRP instances for our experiments.
For each CVRP instance, we take 10 samples.

We implemented both the standard QAOA and our QAOAnsatz.
In OAQAnsatz, the one-hot constraint in Eq.~\eqref{eq:total_alim} is solved using a mixer Hamiltonian, while in QAOA, it is solved using a penalty method.
For simplicity, we set the $\lambda_1$ and $\lambda_2$ as $1$ in Eq.~\eqref{eq:total_alim}.
Due to the size of the instance we are testing, $1$ is enough to keep the constraint.
We used the COBYLA method to optimize the variational parameters~$\bm{\beta},\bm{\gamma}$ for both QAOA and QAOAnsatz and took $1000$ shots for each circuit run.
We used Qamomile\footnote{https://github.com/Jij-Inc/Qamomile}, a quantum optimization library, to construct the Hamiltonian using the ALiM technique for the given subproblem.
The algorithm implementation was carried out using QURI Parts\footnote{https://github.com/QunaSys/quri-parts}, an open-source library designed for experimenting with quantum algorithms.
We calculate reference solution produced by Google OR-Tools.\footnote{\url{https://developers.google.com/optimization}}
For the restricted master problem, we use the CBC solver~\cite{cbc_solver}.

First, we compare the convergence behavior of QAOA and QAOAnsatz for the CVRP instance with four customers and $T=4$.
In Fig.~\ref{fig:compare}, we plot the objective value in Eq.~\eqref{Eq:cg} at each CG iteration step.
The objective value indicates the gap between the current and optimal solution.
Moreover, the objective value equals $0$, indicating that the subproblem cost has reached its minimum possible value for these instances.
The results show that QAOA struggles to reach the optimal solution until 10 steps.
In contrast, QAOAnsatz demonstrates better performance by leveraging its feasibility-preserving design.
It converged after four steps in most of the instances.
These results indicate the design of the mixer operator is important to increase the performance of the CG method.

\begin{figure}[t]
\centering
\includegraphics[width=\linewidth]{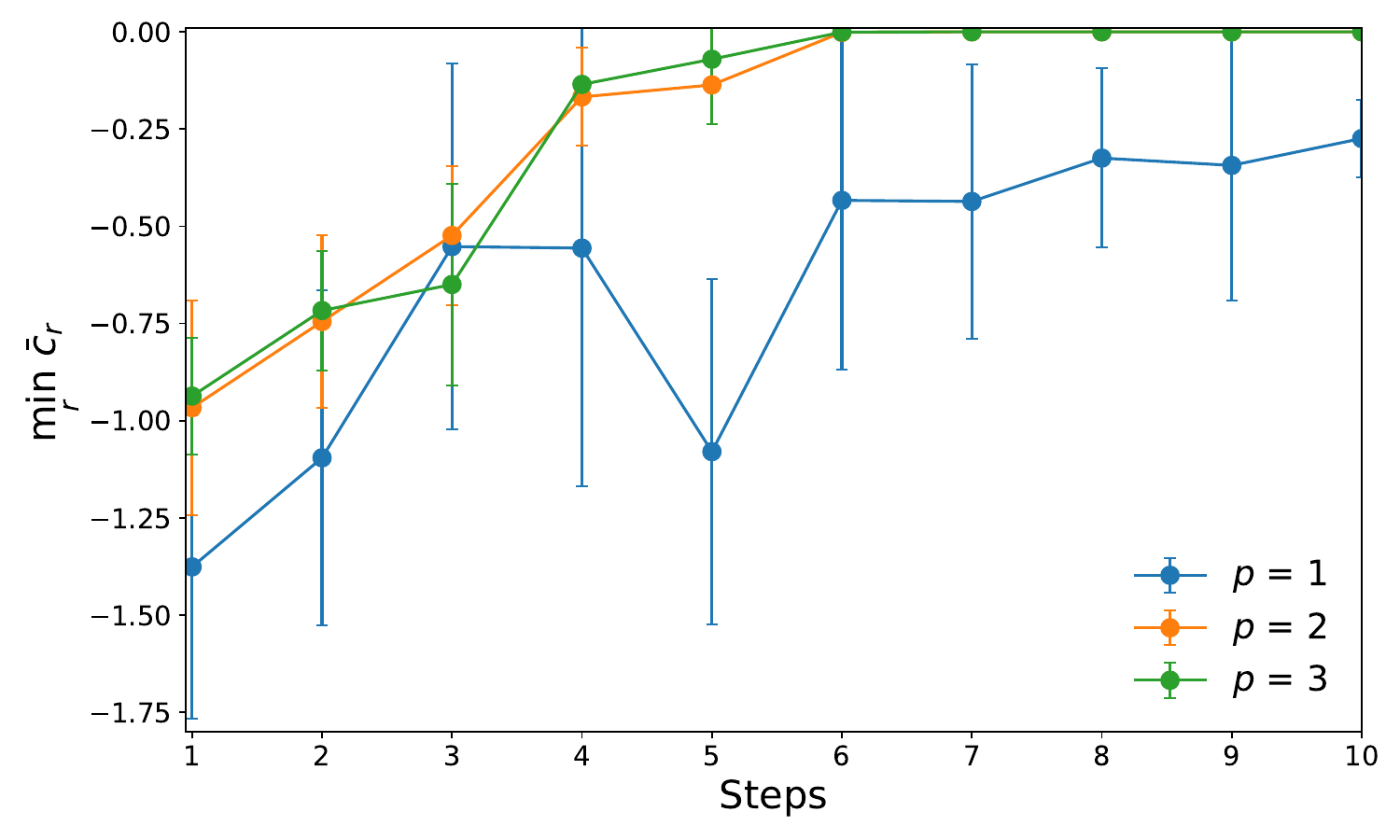}
\caption{
$p$-dependence of the convergence speed of the subproblem objective value, $\min_r\bar{c}_r$ for five customer problem.
$p \geq 2$ results show better converence than the $p=1$ case.
}
\label{fig:P_impact}
\end{figure}

Next, we investigate the influence of the number of the QAOAnsatz layer $p$ on the convergence behavior of our algorithm.
Figure~\ref{fig:P_impact} shows the $p$-dependence of the convergence behavior for five customers.
It shows $p\geq 2$ cases converge much faster than $p=1$ cases.
We set $p$ larger than $2$, which enables the algorithm to explore more potential routes, including those that might be overlooked at $p=1$.
This wider exploration leads to faster convergence. 
However, it also increases the computational cost because each additional QAOA layer increases the circuit depth and the total run time.
Thus, we must balance the algorithm performance and computational cost. 
The convergence speeds of $p=2$ and $p=3$ are almost the same, so in the following experiments, we set $p=2$.

\begin{figure}[t]
\centering
\includegraphics[width=\linewidth]{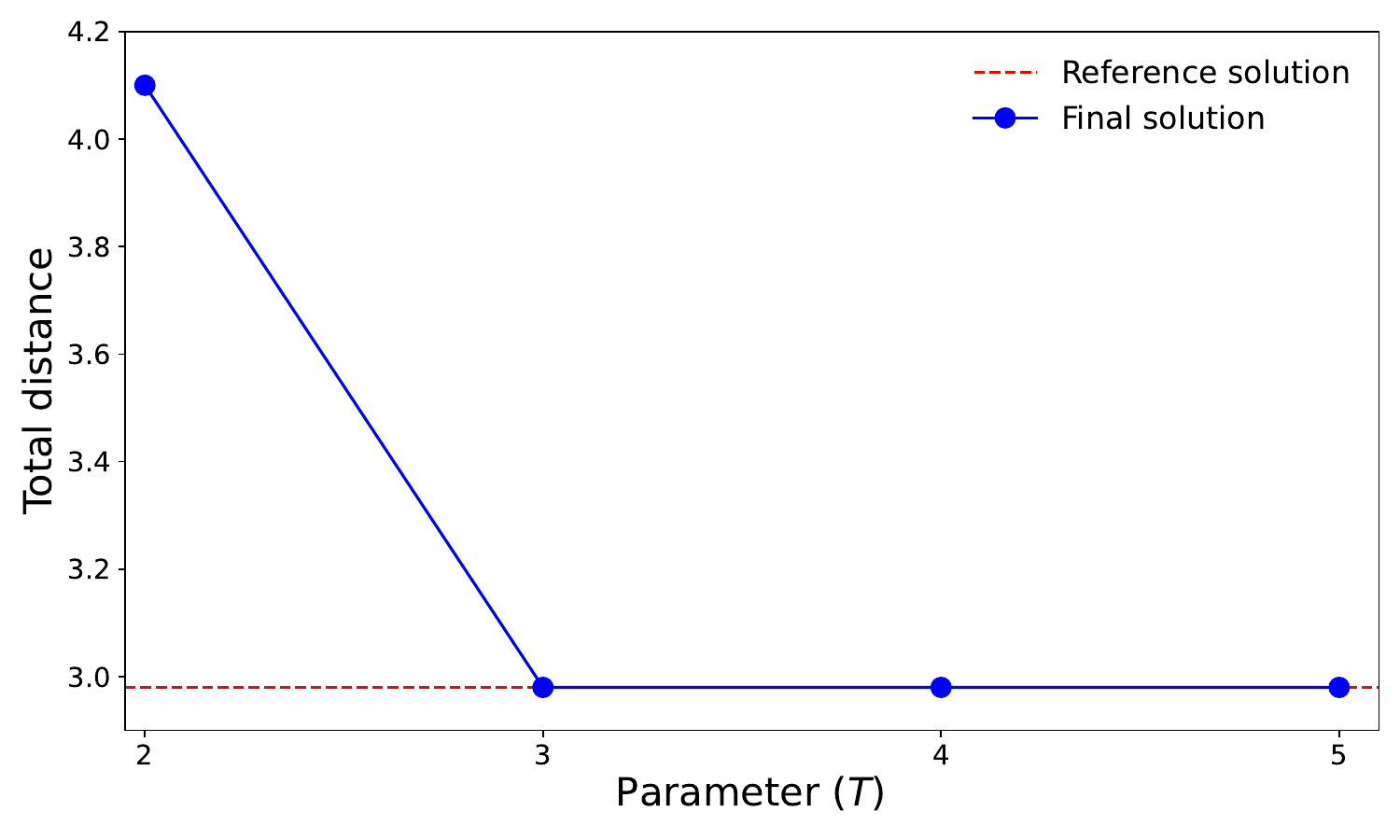}
\caption{
Dependence of CVRP solution distances on $T$ for instances with five customers using the QAOAnsatz.
We can obtain even if we set the smaller $T$ compared to $N$.
}
\label{fig:T_impact}
\end{figure}
\begin{figure*}[t]
\centering
\includegraphics[width=\linewidth]{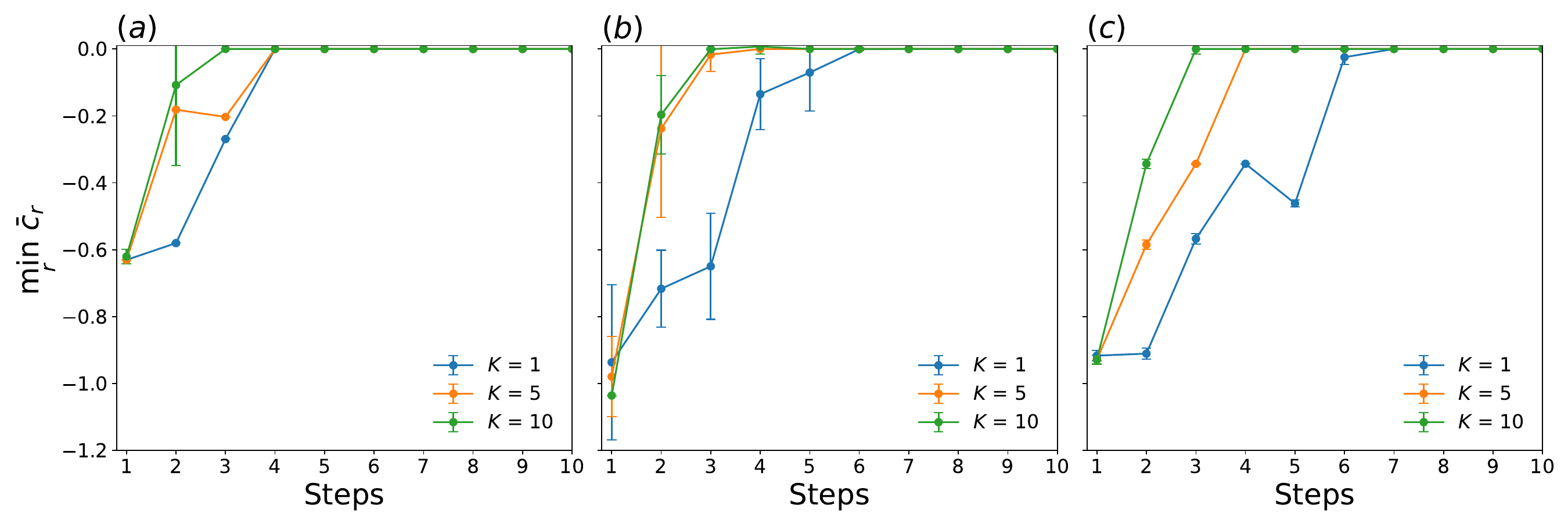}
\caption{
The subproblem objective values, $\min_r\bar{c}_r$, under varying $K$ using the QAOAnsatz method.
Each plot corresponds to the results for (a)~four customers, (b)~five customers, and (c)~six customers. 
Each line corresponds to a different $K$, with error bars indicating the standard deviation.
In all results, increasing $K$ makes better convergence.
}
\label{fig:Num_customers=6T=4}
\end{figure*}

In our CVRP subproblems, the time parameter $T$ need not equal the total number of customers $N$ since a vehicle can complete its route 
before serving all customers. 
In our experiments, we considered a problem instance with five customers.
We varied $T$ from $2$ to $5$ to evaluate its impact on the final solution's total distance traveled by all vehicles under our method compared to the reference solution's total distance.
The results, illustrated in Fig.~\ref{fig:T_impact}, reveal that even when $T$ is much smaller than $N$, it converges to the optimal solution.
It is worth noting that this property is highly dependent on the setting of $W$, and if $W$ is large enough, it is clear that a single vehicle can deliver to all customers, so $T$ cannot be made smaller than $N$. 
However, in realistic settings where multiple vehicles are required, it is expected that the problem size can be reduced.

Finally, we consider that the input route number~$K$ is used to add to the route set~$\mathcal{R}_k$.
Unlike traditional methods, which often derive a single optimal solution at each step, our approach generates multiple candidate solutions.
By adding the optimal solution to the subproblem and solutions in the vicinity of the optimal solution to $\mathcal{R}_k$, we can expect faster convergence.
We analyze the effect of the input route number, $K$, on convergence speed across four to six customers' CVRP instances with $T=4$.
Fig.~\ref{fig:Num_customers=6T=4} shows the subproblem objective values against the number of steps for each instance by choosing $K=1, 5,$ and $10$.
Fig.~\ref{fig:Num_customers=6T=4} (a) to (c) are the results for four to six customers, respectively.
We can see that the $K=1$ results are the slowest to converge in all problem sizes.
Moreover, the result with $K=10$ converges the fastest.
However, the benefit of selecting a larger $K$ depends on the problem size.
In the case of four or five customers, the results show that the convergence speed is only slightly faster for $K=10$ than for $K=1$, but in the case of 6 customers, we can see a significant improvement in convergence speed.
As $K$ increases, it takes longer to solve the restricted master problem, so it is necessary to set $K$ appropriately depending on $N$.

\section{summary}
\label{sec:summary}

In this study, we introduced a hybrid quantum-classical approach to solving the CVRP by combining the CG method with the QAOAnsatz.
CG is a classical decomposition technique often used in large-scale linear or integer programming problems like the CVRP.
Instead of formulating and solving the entire problem at once, CG iteratively finds the best combination of the routes from the route set and adds a new route by solving the subproblem.
This process continues until no improving route can be found.
By leveraging the QAOAnsatz within the CG framework, our hybrid approach addresses the challenge of finding optimal solutions within the feasible solution space
This challenge is that the original QAOA often fails to achieve due to its reliance on solution relaxation.

To demonstrate the effectiveness of our method, we compared the performance of QAOA and QAOAnsatz on CVRP instances with four customers.
As shown in Fig.~\ref{fig:compare}, QAOAnsatz consistently achieved faster convergence to the optimal solution.
The improved performance of QAOAnsatz is attributed to its ability to restrict the search space to feasible regions through the use of tailored mixer Hamiltonians.
In Fig.~\ref{fig:P_impact}, we show how the different layers $p$ affect the convergence.
The result shows that larger $p$ converges faster than $p=1$.
In Fig.~\ref{fig:T_impact}, we examine the influence of varying the time parameter, $T$.
The experiments showed that choosing smaller $T$ does not violate the optimality of the problem.
This result indicates that we can set a smaller $T$, which reduces the required number of qubits.
Moreover, Fig.~\ref{fig:Num_customers=6T=4} illustrates the impact of increasing the input route number, $K$, added to the route set of the restricted master problem on convergence speed.
We showed that choosing larger $K$ improves the convergence speed.

Despite the promising results, this study primarily focused on showing the effectiveness of the CG method and the influence of some parameters, such as the number of layers and the number of inputs route $K$.
In this study, we only solved the relatively minor problems.
Expanding the approach to larger CVRP instances is one of the most important aspects of future work.
There are several ways to improve the solution quality for larger instances.
Combining classical postprocess is one of the methods to improve the solution qualities~\cite{kanai2024annealing}.
Another direction is the combining condition encoding method to solve the subproblems~\cite{Xie2024-sb}.
These methods enhance the precision of the subproblem solutions.

While we apply the CG method with the QAOAnsatz to CVRP in this study, it is essential to highlight that our proposed method applies to a broader range of problems, such as the cutting stock problem~\cite{gilmore1961linear}.
Applying our proposed algorithm to these problems is promising future work.
Integrating quantum algorithms into classical optimization workflows provides a promising direction for leveraging quantum computing capabilities in practical applications.

\printbibliography[title=References]

\end{document}